 \documentclass[11pt,preprint]{aastex} 

\shorttitle{Quasi-Coherent Oscillations}
\shortauthors{Mauche \& Robinson}

 \slugcomment{Accepted for publication in 
             {\it Ap.J.\/} August 2, 2001}

\newcommand\Mdot  {\dot{M}}
\newcommand\Mwd   {M_{\rm wd}}
\newcommand\Rwd   {R_{\rm wd}}
\newcommand\Msun {{\rm M_{\odot}}}
\newcommand\lax{{\lower0.75ex\hbox{ $<$ }\atop\raise0.5ex\hbox{ $\sim$ }}}
\newcommand\gax{{\lower0.75ex\hbox{ $>$ }\atop\raise0.5ex\hbox{ $\sim$ }}}

\begin{document}

\title{First Simultaneous Optical and EUV Observations \\ 
       of the Quasi-Coherent Oscillations of SS~Cygni}

\author{Christopher W.\ Mauche}
\affil{Lawrence Livermore National Laboratory,
       L-43, 7000 East Avenue, Livermore, CA 94550; \\
       mauche@cygnus.llnl.gov}
\and
\author{Edward L.~Robinson}
\affil{Department of Astronomy, University of Texas,
       Austin, TX 78712; \\
       elr@astro.as.utexas.edu}

\clearpage 


\begin{abstract}

Using EUV photometry obtained with the {\it Extreme Ultraviolet
Explorer\/} ({\it EUVE\/}) satellite and $UBVR$ optical photometry
obtained with the 2.7-m telescope at McDonald Observatory, we have
detected quasi-coherent oscillations (so-called ``dwarf nova
oscillations'') in the EUV and optical flux of the dwarf nova SS~Cygni
during its 1996 October outburst. There are two new results from these
observations. First, we have for the first time observed ``frequency
doubling:'' during the rising branch of the outburst, the period of
the EUV oscillation was observed to {\it jump\/} from 6.59~s to 2.91~s.
Second, we have for the first time observed quasi-coherent oscillations
{\it simultaneously\/} in the optical and EUV. We find that the period
and phase of the oscillations are the same in the two wavebands,
finally confirming the long-held assumption that the periods of the
optical and EUV/soft X-ray oscillations of dwarf novae are equal. The
$UBV$ oscillations can be simply the Rayleigh-Jeans tail of the EUV
oscillations if the boundary layer temperature $kT_{\rm bb}\lax 15$~eV
and hence the luminosity $L_{\rm bb}\gax 1.2\times 10^{34}\, (d/{\rm
75~pc})^2~\rm erg~s^{-1}$ (comparable to that of the accretion disk).
Otherwise, the lack of a phase delay between the EUV and optical
oscillations requires that the optical reprocessing site lies within
the inner third of the accretion disk. This is strikingly different from
other cataclysmic variables, where much or all of the disk contributes
to the optical oscillations.

\end{abstract}


\clearpage 
 

\section{Introduction}

Rapid periodic oscillations are observed in the optical flux of high
accretion rate (``high-$\Mdot $'') cataclysmic variables (nova-like
variables and dwarf novae in outburst) \citep{pat81, war95a, war95b}.
These oscillations have high coherence ($Q\approx 10^4$--$10^6$),
short periods ($P\approx 7$--40~s), low amplitudes ($A\lax 0.5$\%), and
are sinusoidal to within the limits of measurement. They are referred
to as ``dwarf nova oscillations'' (DNOs) to distinguish them from the
apparently distinct longer period, low coherence ($Q\approx 1$--10)
quasi-periodic oscillations (QPOs) of high-$\Mdot $ cataclysmic
variables, and the longer period, high coherence ($Q\approx
10^{10}$--$10^{12}$) oscillations of DQ~Her stars. DNOs appear on the
rising branch of the dwarf nova outburst, typically persist through
maximum, and disappear on the declining branch of the outburst. The
period of the oscillation decreases on the rising branch and increases
on the declining branch, but because the period reaches minimum about
one day after maximum optical flux, dwarf novae describe a loop in a
plot of oscillation period versus optical flux.

The dwarf nova SS~Cygni routinely exhibits DNOs during outburst. Optical
oscillations have been detected at various times with periods ranging 
from 7.3~s to 10.9~s \citep{pat78, hor80, hil81, pat81}. In the soft
X-ray and EUV wavebands, quasi-coherent oscillations have been detected in
{\it HEAO 1\/} LED~1 data at periods of 9~s and 11~s \citep{cor80, cor84},
{\it EXOSAT\/} LE  data at periods between 7.4~s and 10.4~s \citep{jon92},
{\it EUVE\/}   DS  data at periods between 2.9~s and  9.3~s \citep{mau96,
mau98}, and
{\it ROSAT\/}  HRI data at periods between 2.8~s and  2.9~s \citep{tes97}.
\citet{mau96} showed that the period of the EUV oscillations of SS~Cyg
is a single-valued function of the EUV flux (hence, by inference, the
mass-accretion rate onto the white dwarf), and explained the loops
observed in plots of oscillation period versus optical flux as the
result of the well-known delay between the rise of the optical and EUV
flux at the beginning of dwarf nova outbursts. While the quasi-coherent
oscillations of SS~Cyg are usually sinusoidal to high degree,
\citet{mau97} pointed out the pronounced distortion of the EUV waveform
at the peak of the 1994 June/July outburst. We present here new
observations in the optical and EUV obtained during the 1996 October
outburst of SS~Cyg.

\section{Observations}

\subsection{EUVE}

As discussed by \citet{whe00}, AAVSO optical, {\it EUVE\/}, and {\it
RXTE\/} observations of SS~Cyg were obtained during a multiwavelength
campaign in 1996 October designed to study the wavelength dependence of
the outbursts of dwarf novae. The {\it EUVE\/} \citep{bow91, bow94}
observations began on $\rm JD-2450000=366.402$ and ended on $\rm JD
-2450000=379.446$. Data are acquired only during satellite night, which
comes around every 95 min and lasted for 23 min (at the beginning of the
observation) to 32 min (at the end of the observation). Valid data are
collected during intervals when various satellite housekeeping monitors
[including detector background and primbsch/deadtime corrections] are
within certain bounds. After discarding numerous short ($\Delta t\le
10$~min) data intervals comprising less than 10\% of the total exposure,
we were left with a net exposure of 208 ks. EUV photometry is provided
both by the deep survey (DS) photometer and short wavelength (SW)
spectrometer, but the count rate is too low and the effective background
is too high to detect oscillations in the SW data. Unfortunately, the DS
photometer was switched off between October 11.37 UT and October 14.70 UT
because the count rate was rising so rapidly on October 11 that the
{\it EUVE\/} Science Planner feared that the DS instrument would be
damaged while the satellite was left unattended over the October 12--13
weekend. We constructed an EUV light curve of the outburst from the
background-subtracted count rates registered by the two instruments,
using a 72--130~\AA \ wavelength cut for the SW spectroscopic data, and
applying an empirically-derived scale factor of 14.93 to the SW count
rates to match the DS count rates. The resulting EUV light curve is
shown by the filled symbols in the upper panel of Figure~1, superposed
on the AAVSO optical light curve shown by the small dots (individual
measurements) and histogram (half-day average). As shown by \citet{mau01},
the EUV light curve {\it lags\/} the optical light curve by $\approx
1{1\over 2}$ days during the rising branch of the outburst, then {\it
leads\/} the optical light curve during the declining branch of the
outburst. The secondary maximum of the EUV light curve at the very end of
the optical outburst appears to be real, and coincides with the recovery
of the hard X-rays flux measured by {\it RXTE\/} \citep{whe00}.

To determine the period of the oscillations of the EUV flux of SS~Cyg, 
for each valid data interval we calculated the power spectra of the 
background-subtracted count rate light curves using 1.024~s bins (the
bin width of the primbsch/deadtime correction table). Individual spectra
typically consist of a spike superposed on a weak background due to
Poisson noise, so in each case we took as the period of the oscillation
the location of the peak of the power spectrum in the interval $\nu=
0.1$--$0.4~\rm Hz$ ($P=2.5$--10~s). The resulting variation of the period
of the EUV oscillation is shown in the lower panel of Figure~1. The
oscillation was first convincingly detected on the rising branch of the
outburst at a period of 7.81~s, fell to 6.59~s over an interval of 4.92~hr
($Q=1.5\times 10^4$), {\it jumped\/} to 2.91~s, and then fell to 2.85~s
over an interval of 4.92~hr ($Q=3.0\times 10^5$) before observations
with the DS were terminated. When DS observations resumed 3.4 days
later during the declining branch of the outburst, the period of the EUV
oscillation was observed to rise from 6.73~s to 8.23~s over an interval
of 2.10 days ($Q=1.2\times 10^5$).

It is clear from the lower panel of Figure~1 that the period of the EUV
oscillation of SS~Cyg anticorrelates with the DS count rate, being long
when the count rate is low and short when the count rate is high. To
quantify this trend, we plot in Figure~2 the log of the period of the
oscillation as a function of the log of the DS count rate. As in the
previous figure, the data fall into two groups: one during the early
rise (distinguished with crosses) and decline of the outburst, the other
during the interval after the frequency of the oscillation had doubled.
The trend during the early rise and decline of the outburst is clearly
the same; fitting a function of the form $P= P_0\, I^{-\alpha}$, where
$I$ is the DS count rate, an unweighted fit to the data gives $P_0=
7.26$~s and $\alpha=0.097$. A similar fit to the data acquired after
the oscillation frequency had doubled gives $P_0=2.99$~s and $\alpha
=0.021$. The first trend is consistent with that observed during
outbursts of SS~Cyg in 1993 August and 1994 June/July \citep{mau96}, but
the trend after the frequency had doubled is clearly distinct: not only
did the oscillation frequency double, it's dependence on the DS count
rate became ``stiffer'' by a factor of $\approx 5$ in the exponent.
SS~Cyg seems to have been doing what it could to avoid oscillating
faster than about 2.8~s. If this is the Keplerian period of material at
the inner edge of the accretion disk, then $P_{\rm Kep}\ge 2\pi (\Rwd
^3/G\Mwd)^{1/2}\approx 2.8$~s, requiring $\Mwd\ge 1.27~\Msun $ (assuming
the \citealt{nau72} white dwarf mass-radius relationship). If instead,
$P_{\rm Kep}\approx 5.6$~s (i.e., the observed 2.8~s period is the first
harmonic of a 5.6~s Keplerian period), then $\Mwd\gax 1.08~\Msun $. The
data of \citet{hes84}, \citet{fri90}, and \citet{mar94} are consistent
with a binary inclination $i\approx 40^\circ $ and white dwarf mass
$\Mwd=0.9$--$1.1~\Msun $, hence favor the second option, but it requires
only a $\approx 10\%$ reduction in the inclination angle to accommodate
the first option.

\subsection{Optical}

In an effort to obtain the first simultaneous optical and EUV/soft X-ray
measurements of dwarf nova oscillations, optical photometry of SS~Cyg
was obtained with the 2.7-m telescope at McDonald Observatory and the
Stiening high-speed photometer on the nights of 1996 October 13, 14, and
15 UT. The Stiening photometer simultaneously measures the flux in four
bandpasses similar to the Johnson $UBVR$ bandpasses (see \citealt{rob95}
for the effective wavelengths and widths of the bandpasses). Fluxes
were calibrated using the standard star BD+28$^\circ$4211, and the time
standard was UTC as given by a GPS receiver located at the dome of the
telescope. The start times of the observations were $\rm JD-2450000=
369.583$, 370.644, and 371.587; the run lengths were 3.14, 1.37, and 3.97
hr respectively; and sample intervals were 0.5~s throughout. SS~Cyg was
observed to fade by $\sim 0.25$ mag between the first and second nights
and by another $\sim 0.30$ mag between the second and third nights,
but the mean flux ratios remained nearly constant from night to night:
the $F_\nu $ flux ratios are $F_U/F_V\approx 1.30$, $F_B/F_V\approx 1.13$,
and $F_R/F_V=0.87$, with $V=9.04$, 9.27, and 9.56 for October 13, 14, and
15 UT, respectively.

A search was made for optical oscillations by calculating the power
spectra of the light curves in the various bandpasses. We found no
detectable periodicities in the light curves from the first night with
an upper limit on the relative amplitude $\Delta F/F<3.0\times 10^{-4}$
for any periodicity between 2.5~s and 10~s. Oscillations were detected
on the second and third nights with periods of 6.58~s and 6.94~s,
respectively. The mean properties of these oscillations are listed in
Table~1. The fluxes in that table should be accurate to a few percent,
and the oscillation amplitudes from the third night also should be
accurate to a few percent, but on the second night the accuracy of the
amplitude measurements are no better than $\sim 20\%$ because the light
curves are weak and contaminated by noise so the oscillation amplitudes
are poorly determined and biased upwards by noise.

The band fluxes $F$, oscillation amplitudes $\Delta F$, and relative
amplitudes $\Delta F/F$ from the third night are plotted in Figure~3,
where it is apparent that the continuum flux of SS~Cyg rises
monotonically from $R$ through $U$, while the absolute and hence the
relative oscillation amplitudes are smallest in $V$. As shown by the
dotted line, the $UBV$ spectral distribution of the oscillation
amplitudes is reasonably consistent with Rayleigh-Jeans, whereas the
spectral distribution of the continuum is much flatter: an unweighted fit
to the $UBV$ measurements of the oscillation amplitudes and continuum
assuming a function of the form $F_\nu \propto\lambda ^{-\alpha }$ yields
$\alpha =1.9$ and 0.57, respectively.

\subsection{Optical and EUV}

To compare the periods of the oscillations detected in the EUV and
optical, we added the points for the optical periods to the lower panel
of Figure~1. The period of the oscillation on the second night is about
what one would expect from an extrapolation of the {\it EUVE\/} data,
but, more importantly, the period of the oscillation on the third night
is consistent with the value measured contemporaneously by {\it EUVE\/}.
To investigate this further, we focused the analysis on that
(unfortunately short) interval when optical and {\it EUVE\/} data were
obtained simultaneously: during $\rm JD-2450000=371.6639$--371.6724 and
371.7297--371.7417; two stretches of strictly simultaneous observations
separated by an {\it EUVE\/} orbit. Given the low amplitude of the
optical oscillations, we calculated power spectra of the various bands
in the encompassing interval $\rm JD-2450000= 371.6639$--371.7417. For
the {\it EUVE\/} data, we calculated the power spectra separately for the
two intervals. These spectra are plotted in the left panels of Figure~4
(where the EUV power spectrum is the average of the two intervals).
In each case (four optical bands, two EUV intervals), the period of the
oscillation is found to be 6.94~s (actually, $6.944\pm 0.004$~s for
the optical channels, $6.94\pm 0.02$~s for the EUV channel; the higher
accuracy for the optical channels is due to longer data interval). To
determine the relative phase of these oscillations, we phase-folded
the data assumed a common period of 6.94~s and a common zero point at
$\rm JD-2450000=371.7$, midway between the two data intervals. Fitting a
sine wave $F +\Delta F\, \sin 2\pi(\phi-\phi_0)$ to each band separately,
we derived the parameters listed in Table~2; the phase-folded light
curves and sinusoidal fits are shown in the right panels of Figure~4.
The primary result from these efforts is that the relative phase of the
oscillation is the same for all the bands. In particular, the phase of
the EUV and optical oscillations are the same within the errors: the
difference is $\Delta\phi _0=0.014\pm 0.038$. The relative amplitudes
derived by these means are higher than derived in the previous section,
but it is to be expected that higher amplitudes will be derived over
short intervals when the oscillation period is more nearly constant. As
before, the oscillation amplitude is highest in $U$, lowest in $V$, and
comparable at an intermediate value in $B$ and $R$.

\section{Summary and Discussion}

We have described EUV and optical photometric observations of SS~Cyg
obtained during its 1996 October outburst. During the rise to outburst,
the period of the EUV oscillation was observed to fall from 7.81~s to
6.59~s over an interval of 4.92~hr, {\it jump\/} to 2.91~s, and then
fall to 2.85~s over an interval of 4.92~hr. During the decline from
outburst, the period of the EUV oscillation was observed to rise from
6.73~s to 8.23~s over an interval of 2.10 days. Optical oscillations
were detected on the second and third nights of observations during the
decline from outburst with periods of 6.58~s and 6.94~s, respectively.
During the times of overlap between the optical and EUV observations on
the third night, the oscillations were found to have the same period and
phase; they differ only in their amplitudes, which are 34\% in the EUV
and 0.05\%--0.1\% in the optical.

The first striking aspect of these observations is the frequency doubling
observed on the rise to outburst. SS~Cyg appears to have undergone a
``phase transition'' at a critical period $P_{\rm c}\lax 6.5$~s, when the
frequency of its oscillation doubled (Fig.~1) and the ``stiffness'' of
its period-intensity relation ($P\propto I^{-\alpha }$) increased by a
factor of $\approx 5$ in the exponent (Fig.~2). Optical oscillations were
detected on the second and third nights of observations at periods above
$P_{\rm c}$, but not on the first night when an extrapolation of the
trend would predict an oscillation period below $P_{\rm c}$. It is
interesting to speculate that the optical oscillations of SS~Cyg (and
possibly other dwarf novae) disappear on the rise to outburst when (if)
the source makes the transition to the higher oscillation frequency and
``stiffer'' period-intensity state, and then reappear on the decline from
outburst when the source reverts back to its normal state. Additional
simultaneous optical and EUV/soft X-ray observations are required to
determine if this is the case. Such observations are also required to
determine if this transition takes place at the same period on the rise
to and decline from outburst, or whether there is a ``hysteresis'' in the
transition. The observational data are consistent with SS~Cyg pulsating at
a fundamental period $P\gax 6.5$~s, then switching to a first harmonic
and stiffening its period-intensity (by inference period-$\Mdot $)
relationship so as to avoid oscillating faster than $P_{\rm min}/2\approx
2.8$~s. This minimum period $P_{\rm min}\approx 5.6$~s is consistent with
the Keplerian period at the inner edge of the accretion disk if, as seems
to be the case observationally, the binary inclination $i\approx 40^\circ
$ and the mass of the white dwarf $\Mwd \approx 1~\Msun $. A secure white
dwarf mass would confirm this interpretation.

The second striking aspect of these observations is the lack of a phase
delay between the EUV and optical oscillations measured simultaneously
on the third night (Fig~4). The relative phase delay $\Delta\phi _0
=0.014\pm 0.038$ for $P=6.94$~s, or $\Delta t=0.10\pm 0.26$~s. The
$3\,\sigma $ upper limit $\Delta t\le 0.88$~s corresponds to a distance
$r =c\, \Delta t\le 2.6\times 10^{10}$~cm. If the EUV oscillation
originates near the white dwarf, and the optical oscillation is formed
by reprocessing of EUV flux in the surface of the accretion disk, the
delays $\Delta t=r\, (1-\sin i\, \cos \varphi )/c$, where the binary
inclination $i\approx 40^\circ $ and $0\le \varphi \le \pi$ is the
azimuthal angle from the line of sight. Then, the distance to the
reprocessing site $r=c\, \Delta t/(1-\sin i\, \cos \varphi )\le 1.6
\times 10^{10}$~cm. To give a sense of scale, this is about 30 white
dwarf radii or one-third the size of the disk. In contrast, eclipse
observations of UX~UMa \citep{nat74, kni98}, Z~Cha \citep{war78}, and
HT~Cas \citep{pat81} indicate that in these high-$\Mdot $ cataclysmic
variables much or all of the disk contributes to the optical oscillations.
The much smaller size of the optical emission region in SS~Cyg is derived
from an application of echo mapping, made possible for the first time by
our strictly simultaneous optical and EUV observations.

The other diagnostic of the optical oscillations is their spectrum,
which is nearly Rayleigh-Jeans in $UBV$ (Fig.~3). Given this result,
it is interesting to ask if the $UBV$ oscillations of SS~Cyg are
simply due to the Rayleigh-Jeans tail of the spectrum responsible for
the EUV oscillations. \cite{mau95} discuss the EUV spectrum of SS~Cyg
and show that it can be parameterized in the 72--130~\AA \ {\it EUVE\/}
SW bandpass by a blackbody absorbed by a column density of neutral
material. Acceptable fits to the spectrum are possible for a wide range
of temperatures $kT_{\rm bb}$, hydrogen column densities $N_{\rm H}$,
and luminosities $L_{\rm bb}$, but the tight correlation between these
parameters significantly constrains the allow region of parameter space.
From Figures 8--10 of Mauche, Raymond, \& Mattei, a reasonable set of
acceptable parameters is as listed in the first three columns of Table~3,
where we have assumed a source distance $d=75$ pc and a fiducial SW
count rate of $0.5~\rm counts~s^{-1}$. Scaling to the SW count rate of
$0.083~\rm counts~s^{-1}$ observed during the interval of overlap on
the third night of observations, the fractional emitting area of these
blackbodies $f=L_{\rm bb}/4\pi\Rwd ^2\,\sigma T_{\rm bb}^4$ are as
listed in the fourth column of Table~3 for an assumed white dwarf radius
$\Rwd =5.8\times 10^8$~cm. With the exception of the coolest model, these
fractional emitting areas are smaller than the value $f=H_{\rm bl}/\Rwd
\sim 3\times 10^{-3}$ expected for a boundary layer with a scale height
$H_{\rm bl}$. The $B$ band flux densities $M_B$ of these models are
as listed in the fifth column of Table~3, and after multiplying by the
EUV oscillation amplitude of 34\% they become the oscillation amplitudes
$\Delta M_B $ listed in the sixth column of Table~3. With an observed
oscillation amplitude $\Delta F_B= 4.4\times 10^{-4}$ Jy (Table~2), the
relative model oscillation amplitudes $\Delta M_B/\Delta F_B$ are as
listed in the seventh column of Table~3. We conclude that a single
source can (within a factor of $\lax 3$) produce both the EUV and $UBV$
oscillations of SS~Cyg if its boundary layer temperature $kT_{\rm bb}
\lax 15$~eV and hence its luminosity $L_{\rm bb}\gax 1.2\times 10^{34}\,
(d/{\rm 75~pc})^2~\rm erg~s^{-1}$. Unfortunately, other data cannot
confirm or exclude this possibility. First, while blackbody fits to {\it
HEAO~1} LED~1 and {\it ROSAT\/} PSPC soft X-ray spectra {\it favor\/}
temperatures $kT_{\rm bb}\approx 20$--30 eV, they are not {\it
inconsistent\/} with temperatures as low as $kT_{\rm bb}\approx 15$~eV
\citep{cor80, pon95}. Second, while the strength of the \ion{He}{2}
$\lambda 4686$ emission line at the peak of the outburst implies a
boundary layer luminosity $L_{\rm bb}\approx 5\times 10^{33}\,
(d/75~{\rm pc})^2~\rm erg~s^{-1}$, hence $kT_{\rm bb}\approx 20$~eV,
the luminosity can be increased to the required value if the fraction of
the boundary layer luminosity intercepted by the disk is decreased from 
$\eta=10\%$ to $\eta\approx 2\%$; such a model has the added charm of
producing the {\it expected\/} boundary layer luminosity $L_{\rm bl}
\approx L_{\rm disk}\approx G\Mwd \Mdot /2\Rwd \approx 3\times 10^{34}\,
(d/{\rm 75~pc})^2~\rm erg~s^{-1}$ (Mauche, Raymond, \& Mattei). The
recent {\it Chandra\/} LETG spectrum of SS~Cyg in outburst will better
constrain the boundary layer parameters, hence allow us to determine if
the boundary layer can produce both the EUV and $UBV$ oscillations.

Either way, we are left to explain the enhancement of the oscillation
amplitude in $R$ over that predicted a by Rayleigh-Jeans spectrum
(Fig.~3). We have no compelling explanation for this datum, but the echo
mapping constraint from above still applies, so the source of the extra
oscillation amplitude in $R$ must lie within the inner third of the
accretion disk. We note that \citet{ste01} recently measured the spectrum
of the optical oscillations of the dwarf nova V2051 Oph in outburst and
found that the oscillation amplitudes of the Balmer and He~I emission
lines were stronger than the continuum by factors of $\lax 5$. Our $R$
bandpass contains the H$\alpha $ emission line, so it interesting to
speculate that the enhanced oscillation amplitude in $R$ might be due to
the larger oscillation amplitude of the H$\alpha $ line flux compared
to the continuum. Unfortunately, it seems unlikely that the needed
factor-of-two enhancement can be produced this way. Also, for this
explanation to work, the higher-order Balmer lines must not significantly
enhance the oscillation amplitude in $B$. Fast optical spectroscopy of
SS~Cyg in outburst is required to determine if this scenario can explain
the enhanced oscillation amplitude in $R$, and, more generally, to
determine if our explanation of the origin of the $UBV$ oscillations is
correct. Clearly, there is much more observational work to be done.

\acknowledgments

The {\it EUVE\/} observations of SS~Cyg could not have been accomplished
without the efforts of the members, staff, and director, J.\ Mattei, of
the American Association of Variable Star Observers; {\it EUVE\/} Deputy
Project Scientist R.\ Oliversen; {\it EUVE\/} Science Planner B.\ Roberts;
the staff of the {\it EUVE\/} Science Operations Center at CEA, and the
Flight Operations Team at Goddard Space Flight Center. We thank the
referee for a number of suggestions which improved the clarity of the
manuscript. C.~W.~M.'s contribution to this work was performed under the
auspices of the U.S.\ Department of Energy by University of California
Lawrence Livermore National Laboratory under contract No.\ W-7405-Eng-48.

\clearpage 


\clearpage 


\begin{deluxetable}{cccccc}
\footnotesize
\tablecaption{Mean Properties of the Optical Oscillations\label{tab1}}
\tablewidth{0pt}
\tablehead{
\colhead{Date}              & \colhead{Period}& \colhead{Band-}& \colhead{Flux}&      \colhead{Amplitude}&        \colhead{Relative Amplitude}\\
\colhead{($\rm JD-2450000$)}& \colhead{(s)}&    \colhead{pass}&  \colhead{$F$ (Jy)}& \colhead{$\Delta F$ (Jy)}& \colhead{$\Delta F/F$}
}
\startdata
\hbox to 1.5in{370.644--370.701\leaders\hbox to 0.5em{\hss.\hss}\hfill}&  6.58&  $U$& 0.94& $2.6\times 10^{-4}$& $2.7\times 10^{-4}$\\
                &      &  $B$& 0.81& $1.1\times 10^{-4}$& $1.3\times 10^{-4}$\\
                &      &  $V$& 0.72& $0.8\times 10^{-4}$& $1.1\times 10^{-4}$\\
                &      &  $R$& 0.63& $1.0\times 10^{-4}$& $1.6\times 10^{-4}$\\
\hbox to 1.5in{371.586--371.752\leaders\hbox to 0.5em{\hss.\hss}\hfill}&  6.94&  $U$& 0.70& $3.7\times 10^{-4}$& $5.3\times 10^{-4}$\\
                &      &  $B$& 0.62& $2.3\times 10^{-4}$& $3.7\times 10^{-4}$\\
                &      &  $V$& 0.55& $1.6\times 10^{-4}$& $2.9\times 10^{-4}$\\
                &      &  $R$& 0.48& $1.9\times 10^{-4}$& $3.9\times 10^{-4}$\\
\enddata
\end{deluxetable}

\clearpage 

\begin{deluxetable}{ccccc}
\footnotesize
\tablecaption{Parameters of the Simultaneous EUV and Optical Oscillations\label{tab2}}
\tablewidth{0pt}
\tablehead{
\colhead{Band-}& \colhead{Flux}&                 \colhead{Amplitude}&                   \colhead{Relative Amplitude}& \colhead{Phase}\\
\colhead{pass}&  \colhead{$F$\tablenotemark{a}}& \colhead{$\Delta F$\tablenotemark{a}}& \colhead{$\Delta F/F$}&       \colhead{$\phi _0$}
}
\startdata
\hbox to 0.7in{EUV\leaders\hbox to 0.5em{\hss.\hss}\hfill}& 1.30& $0.45\pm 0.04$              & $0.34\pm 0.03$                         & $0.561\pm 0.014$\\
\hbox to 0.7in{$U$\leaders\hbox to 0.5em{\hss.\hss}\hfill}& 0.71& $(7.7\pm 0.6)\times 10^{-4}$& $(          10.8\pm 0.9)\times 10^{-4}$& $0.577\pm 0.013$\\
\hbox to 0.7in{$B$\leaders\hbox to 0.5em{\hss.\hss}\hfill}& 0.62& $(4.4\pm 0.4)\times 10^{-4}$& $(\phantom{0}7.1\pm 0.6)\times 10^{-4}$& $0.578\pm 0.014$\\
\hbox to 0.7in{$V$\leaders\hbox to 0.5em{\hss.\hss}\hfill}& 0.56& $(2.8\pm 0.4)\times 10^{-4}$& $(\phantom{0}5.0\pm 0.7)\times 10^{-4}$& $0.575\pm 0.022$\\
\hbox to 0.7in{$R$\leaders\hbox to 0.5em{\hss.\hss}\hfill}& 0.48& $(3.4\pm 0.4)\times 10^{-4}$& $(\phantom{0}7.1\pm 0.9)\times 10^{-4}$& $0.567\pm 0.019$\\
\enddata
\tablenotetext{a}{Units: DS counts $\rm s^{-1}$ for the EUV band, Jy for the optical bands}
\end{deluxetable}

\clearpage 

\begin{deluxetable}{ccccccc}
\footnotesize
\tablecaption{Blackbody Model Parameters\label{tab3}}
\tablewidth{0pt}
\tablehead{
\colhead{$kT$}& \colhead{$N_{\rm H}$}&     \colhead{$L_{\rm bb}$}&       \colhead{}&    \colhead{Flux}&       \colhead{Amplitude}&         \colhead{Relative Amplitude}\\
\colhead{(eV)}& \colhead{($\rm cm^{-2}$)}& \colhead{($\rm erg~s^{-1}$)}& \colhead{$f$}& \colhead{$M_B$ (Jy)}& \colhead{$\Delta M_B$ (Jy)}& \colhead{$\Delta M_B/\Delta F_B$}
}
\startdata
\hbox to 0.5in{15\leaders\hbox to 0.5em{\hss.\hss}\hfill}& $9.5\times 10^{19}$& $1.2\times 10^{34}$& $9.3\times 10^{-3}$& $4.3\times 10^{-4}$& $1.5\times 10^{-4}$& $3.3\times 10^{-1}$\\
\hbox to 0.5in{20\leaders\hbox to 0.5em{\hss.\hss}\hfill}& $7.0\times 10^{19}$& $2.2\times 10^{33}$& $5.2\times 10^{-4}$& $3.3\times 10^{-5}$& $1.1\times 10^{-5}$& $2.5\times 10^{-2}$\\
\hbox to 0.5in{25\leaders\hbox to 0.5em{\hss.\hss}\hfill}& $5.4\times 10^{19}$& $8.5\times 10^{32}$& $8.3\times 10^{-5}$& $6.6\times 10^{-6}$& $2.2\times 10^{-6}$& $5.1\times 10^{-3}$\\
\hbox to 0.5in{30\leaders\hbox to 0.5em{\hss.\hss}\hfill}& $4.4\times 10^{19}$& $4.7\times 10^{32}$& $2.2\times 10^{-5}$& $2.1\times 10^{-6}$& $7.2\times 10^{-7}$& $1.6\times 10^{-3}$\\
\hbox to 0.5in{35\leaders\hbox to 0.5em{\hss.\hss}\hfill}& $3.7\times 10^{19}$& $4.0\times 10^{32}$& $1.0\times 10^{-5}$& $1.2\times 10^{-6}$& $3.9\times 10^{-7}$& $8.9\times 10^{-4}$
\enddata
\end{deluxetable}

\clearpage 


\begin{figure}
\figurenum{1}
\epsscale{0.6}
\plotone{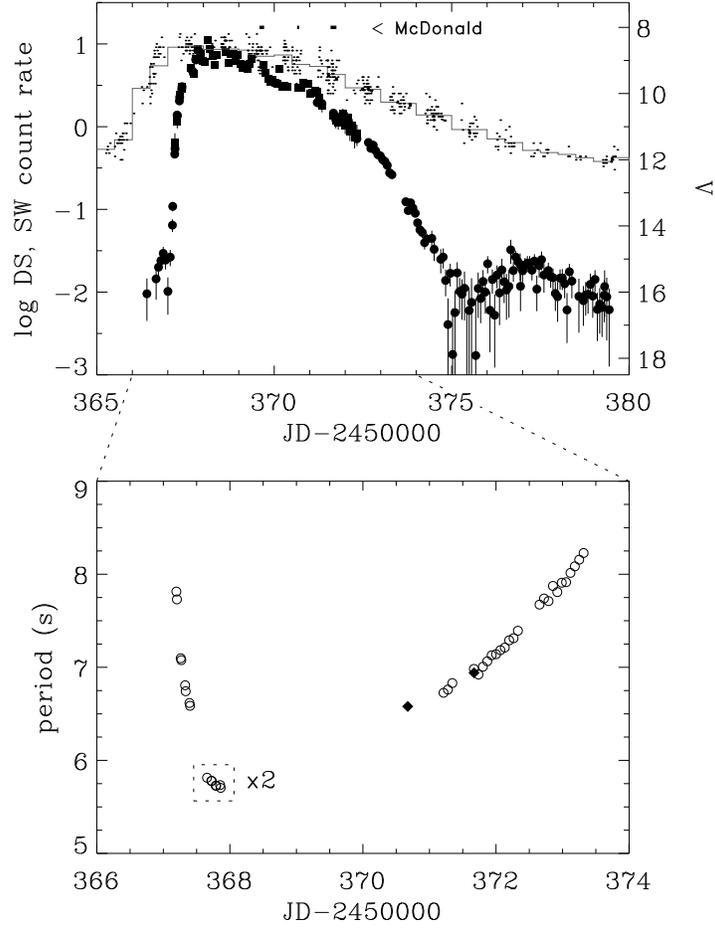}
\caption{{\it Upper panel\/}: {\it EUVE\/} and AAVSO optical light
curves of the 1996 October outburst of SS~Cyg. DS and SW measurements
are shown respectively by the filled circles and squares with error
bars; individual AAVSO measurements are shown by the small dots; 0.5 
day mean optical light curve is shown by the histogram. Intervals of
observations at McDonald Observatory are indicated by the thick bars.
{\it Lower panel\/}: Oscillation period versus time. {\it EUVE\/} DS and
McDonald Observatory optical measurements are shown by the open circles
and filled diamonds, respectively. Points enclosed by the dotted box are
plotted at twice the observed periods.}
\end{figure}

\begin{figure}
\figurenum{2}
\epsscale{0.525}
\plotone{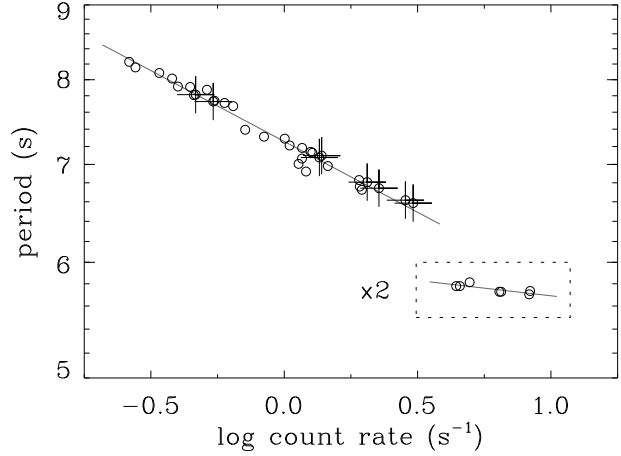}
\caption{Period of the EUV oscillation as a function of DS count rate.
Points on the rising branch of the outburst are distinguished with
crosses. Grey lines are the unweighted fits to the data: $P= 7.26\,
I^{-0.097}$~s and $P=2.99\, I^{-0.021}$~s. Points enclosed by the dotted
box are plotted at twice the observed periods.}
\end{figure}

\begin{figure}
\figurenum{3}
\epsscale{0.579}
\plotone{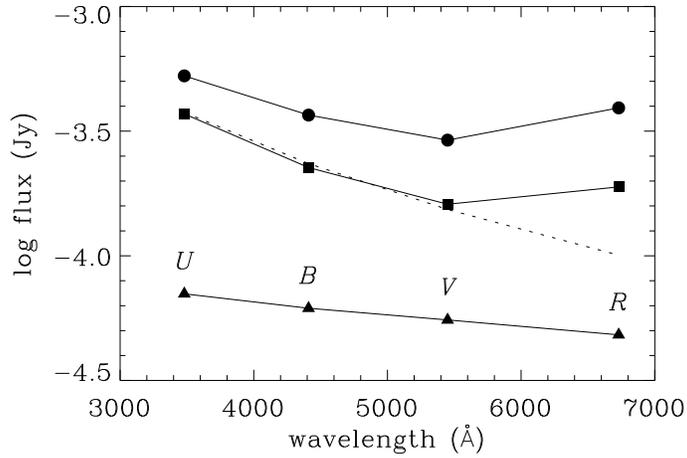}
\caption{Optical flux ($F/10^4$, {\it filled triangles\/}), oscillation
amplitude ($\Delta F$, {\it filled squares\/}), and relative oscillation
amplitude ($\Delta F/F$, {\it filled circles\/}) for 1996 October 15 UT.
Dotted line is a Rayleigh-Jeans spectrum normalized to the $UBV$ fluxes
of the oscillation amplitude.}
\end{figure}

\begin{figure}
\figurenum{4}
\epsscale{1.0}
\plotone{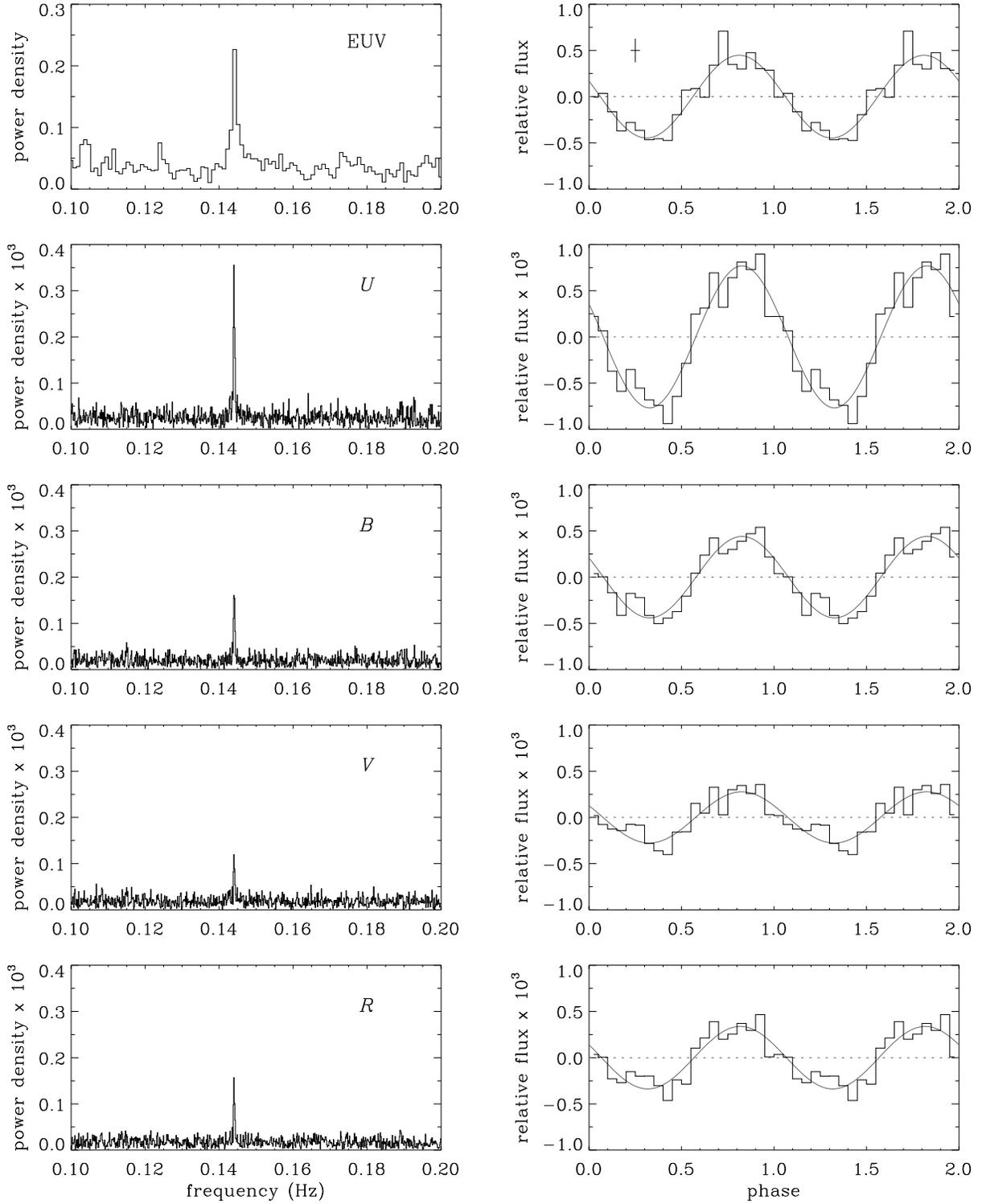}
\caption{Power density ({\it left panels\/}) and phase-folded light 
curves ({\it right panels\/}) of SS~Cyg in the EUV and optical $U$, $B$,
$V$, and $R$ for the intervals of overlap on 1996 October 15 UT. Typical
error bar for the EUV light curve is shown by the cross, and the
sinusoidal fits are the grey curves.}
\end{figure}


\begin{thebibliography}{} 

\bibitem[Bowyer et~al.(1994)]{bow94}
         Bowyer, S., et~al. 1994, \apjs , 93, 569
\bibitem[Bowyer \& Malina(1991)]{bow91}
         Bowyer, S., \& Malina, R.~F. 1991, in Extreme Ultraviolet
         Astronomy, ed.\ R.~F.\ Malina \& S.~Bowyer (New York: 
         Pergamon), 397
\bibitem[C\'ordova et~al.(1984)]{cor84}
         C\'ordova, F.~A., Chester, T.~J., Mason, K.~O., Kahn, S.~M.,
         \& Garmire, G.~P. 1984, \apj , 278, 739
\bibitem[C\'ordova et~al.(1980)]{cor80}
         C\'ordova, F.~A., Chester, T.~J., Tuohy, I.~R., \& Garmire, 
         G.~P. 1980, \apj , 235, 163
\bibitem[Friend et al.(1990)]{fri90}
         Friend, M~T., Martin, J.~S., Smith, R.~C., \& Jones, D.~H.~P.
         1990, \mnras , 246, 654
\bibitem[Hessman et al.(1984)]{hes84}
         Hessman, F.~V., Robinson, E.~L., Nather, R.~E., \& Zhang, 
         E.-H. 1984, \apj , 286, 747
\bibitem[Hildebrand, Spillar, \& Stiening(1981)]{hil81} 
         Hildebrand, R.~H., Spillar, E.~J., \& Stiening, R.~F. 1981,
         \apj , 243, 223
\bibitem[Horne \& Gomer(1980)]{hor80}
         Horne, K., \& Gomer, R. 1980, \apj , 237, 845
\bibitem[Jones \& Watson(1992)]{jon92}
         Jones, M.~H., \& Watson, M.~G. 1992, \mnras , 257, 633
\bibitem[Knigge et al.(1998)]{kni98}
         Knigge, C., Drake, N., Long, K.~S., Wade, R.~A., Horne, K.,
         \& Baptista, R. 1998, \apj , 499, 429
\bibitem[Mart\' \i nez-Pais et al.(1994)]{mar94}
         Mart\' \i nez-Pais, I.~G., Giovannelli, F., Rossi, C., \&
         Gaudenzi, S. 1994, \aap , 291, 455
\bibitem[Mauche(1996)]{mau96}
         Mauche, C.~W. 1996, \apj , 463, L87
\bibitem[Mauche(1997)]{mau97}
         Mauche, C.~W. 1997, in Accretion Phenomena and Related 
         Outflows, ed.\ D.~T.\ Wickramasinghe, L.~Ferrario, \&
         G.~V.\ Bicknell (San Francisco: ASP), 251
\bibitem[Mauche(1998)]{mau98}
         Mauche, C.~W. 1998, in Wild Stars in the Old West, ed.\
         S.~Howell, E.~Kuulkers, C.~Woodward (San Francisco: ASP), 113
\bibitem[Mauche, Mattei, \& Bateson(2001)]{mau01}
         Mauche, C.~W., Mattei, J.~A., \& Bateson, F.~M. 
         2001, in Evolution of Binary and Multiple Stars,  
         ed.\ P.~Podsiadlowski, et al.\ (San Francisco: ASP), 367
\bibitem[Mauche, Raymond, \& Mattei(1995)]{mau95}
         Mauche, C.~W., Raymond, J.~C., \& Mattei, J.~A.  1995, \apj ,
         446, 842
\bibitem[Nather \& Robinson(1974)]{nat74}
         Nather, R.~E., \& Robinson, E.~L. 1974, \apj , 190, 637
\bibitem[Nauenberg(1972)]{nau72}
         Nauenberg, M. 1972, \mnras , 254, 493
\bibitem[Patterson(1981)]{pat81}
         Patterson, J. 1981, \apjs , 45, 517
\bibitem[Patterson, Robinson, \& Kiplinger(1978)]{pat78}
         Patterson, J., Robinson, E.~L., \& Kiplinger, A.~L. 1978,
         \apj , 226, L137
\bibitem[Ponman et al.(1995)]{pon95}
         Ponman, T.~J., et al. 1995, \mnras , 276, 495
\bibitem[Robinson et al.(1995)]{rob95}
         Robinson, E.~L., et al. 1995, \apj , 438, 908
\bibitem[Steeghs et al.(2001)]{ste01}
         Steeghs, D., O'Brien, K., Horne, K., Gomer, R., \& Oke, J.~B.
         2001, \aap , 323, 484
\bibitem[van Teeseling(1997)]{tes97}
         van Teeseling, A. 1997, \aap , 324, L73
\bibitem[Warner(1995a)]{war95a}
         Warner, B. 1995a, Cataclysmic Variable Stars (Cambridge: CUP)
\bibitem[Warner(1995b)]{war95b}
         Warner, B. 1995b, in Cape Workshop on Magnetic Cataclysmic
         Variables, ed.\ D.~A.~H.\ Buckley \& B.~Warner (San Francisco:
         ASP), 343
\bibitem[Warner \& Brickhill(1978)]{war78}
         Warner, B., \& Brickhill, A.~J. 1978, \mnras , 182, 777
\bibitem[Wheatley, Mauche, \& Mattei(2000)]{whe00}
         Wheatley, P.~J., Mauche, C.~W., \& Mattei, J.~A. 2000, New
         Astro.\ Rev., 44, P33
\end{thebibliography}
\end{document}